\documentclass[english,aps,prstper,reprint,showpacs,titlepage,longbibliography]{revtex4-2}   

\usepackage[T1]{fontenc}	
\usepackage[latin9]{inputenc}	
\usepackage{geometry}		
\usepackage{graphicx}
\usepackage[above,below]{placeins}	
\usepackage{times}
\usepackage{enumitem} 
\usepackage{multirow}

\usepackage{hyperref}  
\hypersetup{colorlinks=true,urlcolor=blue,citecolor=blue,linkcolor=blue}   
\usepackage{enumitem}          
\setlist{nosep}                 

\begin{document}

\begin{titlepage}

  \title{
  Exploring the contributions of self-efficacy and test anxiety to gender differences in assessments}
  \author{Jared B. Stang}
  \author{Emily Altiere}
  \author{Joss Ives}
  \affiliation{Dept. of Physics \& Astronomy, University of British Columbia, 6224 Agricultural Road, Vancouver, BC V6T 1Z1} 
  \author{Patrick J. Dubois}
  \affiliation{Dept. of Psychology, University of British Columbia, 2136 West Mall, Vancouver, BC V6T 1Z4} 


  \begin{abstract}
  The observed performance difference between women and men on assessments in physics---the ``gender gap''---is a significant and persistent inequity which has broad implications for the participation of women in physics. Research also shows that gender-based inequities extend to affective measures, such as self-efficacy. In this exploratory study, we report on gender disparities in self-efficacy and test anxiety and their relationship to assessment scores in our active-learning introductory physics course. Overall, gender-based differences in favour of men are observed in all our measures, with women having lower scores on measures associated with success (self-efficacy and assessment scores) and a higher score on a possibly detrimental affective factor (test anxiety). Using a multiple regression model-selection process to explore which measures may explain end-of-course Force Concept Inventory (FCI) and final exam scores, we find that the best fitting models include FCI pretest and self-efficacy as predictors, but do not include test anxiety.
  \clearpage\end{abstract}

  \maketitle
\end{titlepage}

\section{\label{sec:intro}Introduction}

There is a significant and persistent disparity in the participation of women and men in science, technology, engineering, and mathematics (STEM) disciplines. In 2016, women made up only 23\% of the Canadian STEM workforce ages 25-64 \cite{Wall2019} and earned only 36\% of Bachelor's degrees in STEM in the United States; within physics, the share of degrees earned by women that year was only 20\% \footnote{See the data from the APS and IPEDS at \url{https://www.aps.org/programs/education/statistics/}.}. A key issue related to this participation disparity is the consistently observed performance difference between women and men on assessments: the ``gender gap'' \cite{Madsen2013,Day2016,Normandeau2017,Henderson2019,Kost2009a,Kost-Smith2010}. Despite the extensive work by the PER community on gender-based performance differences, the issue remains unresolved and the implications continue to disadvantage women in physics courses and beyond.

A parallel developing research area has been on gender-based disparities in affective measures such as self-efficacy. These have been reviewed for STEM overall \cite{Eddy2016,Trujillo2014} and have been studied in the context of physics \cite{Marshman2018a,Lewis2017,Sawtelle2012}. In this report, we focus on gender disparities in self-efficacy and test anxiety. Self-efficacy influences the choices someone makes and the effort they put forth in a task \cite{Bandura1997}, and it's been shown to be related to performance in physics in \cite{Kost-Smith2011,Sawtelle2012,Kalender2020a} and to physics identity \cite{Hazari2010}. Ballen \emph{et al.} \cite{Ballen2017} showed that self-efficacy mediated learning gains for underrepresented minority students in a biology course, identifying self-efficacy as a possible avenue for supporting equity-seeking groups. Test anxiety is a psychological mechanism which may cause an individual to underperform on an assessment; if an entire demographic group experiences relatively high test anxiety, this could explain observed performance differences. Ballen, Salehi, and Cotner \cite{Ballen2017a} found that---again in a biology course---women reported higher test anxiety than men and that test anxiety had a negative correlation with exam performance for women only. If this were true in physics or generally, then interventions or course designs aimed at reducing test anxiety may help to reduce gender-based differences.

Below, we describe a pilot study in our introductory active-learning physics course. Our research questions are: 1. In our context, are there gender disparities in student self-efficacy, test anxiety, Force Concept Inventory (FCI) pre- and post-test, and exam scores? 2. If gender disparities exist in FCI post-test and exam scores, can self-efficacy and/or test anxiety explain some of the difference? Through this work, we aim to develop a better understanding of the factors involved in gender-based inequalities in physics classrooms, with the ultimate goal of improving the culture for all students.

\section{\label{sec:method}Method}

\subsection{Theoretical framework}
We adopt the model presented by Eddy and Brownell in their review of gender disparities across undergraduate STEM disciplines \cite{Eddy2016}. In their model, persistence in STEM results from both ``observables'' ---performance and engagement---and ``unobservables''---psychological factors such as self-efficacy and belonging, and prior preparation. Gender-based disparities in persistence result from inequalities in the ``observables'' and ``unobservables''. In this study, we focus on the relationship of two unobservable factors---self-efficacy and test anxiety---to the inequalities we observe in assessment scores.

Rodriguez \textit{et. al.} \cite{Rodriguez2012a} provide explicit models of equity that we use to frame our results. In an \emph{equity of parity} model, equity is achieved when the equity-seeking group matches the dominant group on the desired outcome. This model is implicitly adopted in work focused on closing achievement gaps, and we take this perspective in Section~\ref{sec:parity} to examine the gender-based differences on our measures. In an \emph{equity of fairness} model, equity is achieved when all groups receive treatment free from bias. In the example of concept inventories, this would mean that all groups experience equal conceptual gains. This model of equity then maintains disparities in outcomes. In our analysis of exam and FCI scores, Section~\ref{sec:fairness}, this perspective guides our interpretation.

Throughout this manuscript, we disaggregate our data along gender lines, and refer to female students and male students, implicitly adopting the binary gender deficit model \cite{Traxler2016}. We do not take this approach because men are the standard to which women should be compared. Rather, in seeking to understand how sexism may impact women in our physics classrooms, we consider male students as a pseudo-control group that does not experience sexism. Further, we acknowledge that gender identity is not binary, and that the categories of ``female'' and ``male'' do not accurately represent the diversity with which individuals experience their gender and with which gender influences their experience.

There are several important limitations of our present work. Firstly, we do not consider how other identity factors, such as race/ethnicity or socio-economic status, or intersections thereof may impact student experiences in our classroom. A more complete description would take these into account. Secondly, we consider only a global measure of self-efficacy, most similar to a general confidence in ability in the course. It has been shown that different types of self-efficacy may matter differently for women and men \cite{Sawtelle2012}, a nuance which is not taken into account here. Finally, a future analysis should also consider instrument fairness in the FCI, which may explain up to 30\% of the gender difference in FCI scores \cite{Henderson2019}.

\subsection{\label{sec:course}Course description}

The data for this study were collected from ``Physics 1,'' an algebra-based introductory physics course at a large, research-intensive Canadian university. Students enroll in Physics 1 if they did not take physics at the senior level in high school. The course has been transformed to an active learning style as part of the Carl Wieman Science Education Initiative \cite{wieman2017improving}. In the semester analyzed here, 779 students completed the course. The students were spread over three lecture sections which each met for three hours per week. Each section was taught by a white male instructor. Students also attended a weekly two hour recitation, consisting primarily of problem solving in groups. According to institutional records, first-year students comprised 78\% percent of the class.

\subsection{\label{sec:data}Data and data collection}

We measured physics preparation with the FCI \cite{Hestenes1992} and assessed learning results with the FCI and the course exam. The FCI was given to students by Teaching Assistants during recitation sessions in the second week of the course (``FCI pretest'') and again in week 12 of the thirteen week course (``FCI post-test''). Students were asked to not skip questions and to avoid guessing, and were told that the inventory will be used to make future improvements to the course. Students were not given any points for completing either the FCI pre- or post-test. The topics of Physics 1 include kinematics, Newton's laws, energy, and energy and heat transfer mechanisms. 
The course exam consisted of a 130-minute solo phase followed by a 30-minute group phase \cite{Wieman2015}; we use only the solo phase exam scores here, since they represent the performance of the individual students. The solo exam consisted of about half multiple-choice conceptual questions and half open response problems. The exam took place approximately two weeks after the last day of classes.

Gender information was acquired from institutional records. Up to and including the year the data was collected, students selected between radio button options for ``male'' and ``female'' to indicate their ``Gender'' when they created a student account as part of their application to this university. 

To assess student self-efficacy and test anxiety, we adapted survey items from the Motivated Strategies for Learning Questionnaire (MSLQ) \cite{Pintrich1991} related to self-efficacy and anxiety. The MSLQ has 8 items comprising their self-efficacy factor and 5 items comprising their anxiety factor. In addition to these, we included new items related to anxiety (similar to those on the MSLQ but with a positive framing), confidence on the course midterm, and a comparison of anxiety in physics to other science courses. The anxiety-related items were added in an attempt to create a more robust factor, while the other added items were of local interest. Students responded to each of the 21 items on a seven-point Likert scale, from ``strongly disagree'' to ``strongly agree''. The survey was administered during recitation sessions in week 10 of the course, two weeks after the midterm exam and after graded midterm papers were returned to students.

\begin{table*}[t]
    \begin{ruledtabular}
    \begin{tabular}{lcccccccccc}
        & \multicolumn{3}{c}{Female students} & \multicolumn{3}{c}{Male students} & & & &\\
        & $N$ & Mean & SE & $N$ & Mean & SE & Difference in means & 95\% CI & $t$ & Cohen's $d$\\ \hline
        Self-efficacy & 319 & -0.18 & 0.052 & 124 & 0.46 & 0.091 & -0.64*** & [-0.85, -0.43] & -6.10 & -0.67 \\
        Test anxiety & 319 & 0.10 & 0.055 & 124 & -0.26 & 0.088 & 0.38*** & [0.17, 0.58] & 3.66 & 0.38\\
        FCI pretest & 495 & 29.4 & 0.63 & 217 & 43.2 & 1.19 & -13.8*** & [-16.5, -11.1] & -10.2 & -0.91\\
        FCI post-test & 438 & 51.9 & 0.82 & 197 & 64.0 & 1.32 & -12.2*** & [-15.21, -9.09] & -7.81 & -0.69 \\
        Exam & 521 & 61.9 & 0.69 & 243 & 64.5 & 1.10 & -2.61* & [-5.16, -0.06] & -2.01 & -0.16
    \end{tabular}
    \end{ruledtabular}
    \caption{\label{tab:diffs}Summary statistics and differences between female students and male students on the various measures. Self-efficacy and test anxiety scores have been standardized; FCI and exam scores are percentages. Standard errors in the group means (SE) and 95\% confidence intervals (95\% CI) for the difference in means are reported. A negative difference in means implies that the average for female students is lower than the average for male students. The $t$-test is a two-tailed test assuming unequal variance. $d=0.20$ represents a small effect size, $d=0.50$ is a medium effect size, and $d=0.80$ is a large effect size \cite{Maher2013a}. *: $p<.05$; **: $p<.01$; ***: $p<.001$.}
\end{table*}

\subsection{\label{sec:survey}Factor structure of self-efficacy and test anxiety survey}

We used an exploratory factor analysis (EFA) to assess the emergent dimensions that we focus on for analysis. We began by checking the suitability of the data for a factor analysis following the process and thresholds described in Knekta \emph{et al.} \cite{Knekta2019}. For the full 21-item survey, we examined possible outliers, checked for univariate and multivariate normality and linearity of relationships, and confirmed that the data were factorable and did not exhibit multicollinearity. Mardia's tests for multivariate skewness and kurtosis revealed deviations from multivariate normality; however, in the factor analysis, we used the principal axis factor estimator, which is both robust to non-normality and appropriate for ordinal data.


We implemented the EFA using an oblique rotation method to allow correlations between the factors. Based on theoretical grounds we expected 2 factors (self-efficacy and test anxiety); the scree plot for our full survey suggested 2 or 3 factors. We first examined the three-factor EFA, including all survey items. However, in addition to our expectation of just two factors, metrics indicated that the three-factor solution was inadequate: Several items had  low primary loadings, cross-loaded between two factors, had low communalities, and/or had high complexities \cite{Knekta2019}. Altogether, this led us to focus on the two-factor solution.


To determine the two-factor solution, we began with the full survey and undertook a stepwise removal of items with low primary loadings, with a strict cutoff of $|0.5|$ in order to have strong factors. At each step, we removed the item with the smallest primary loading and re-ran the EFA until no more items had a primary loading smaller than our cutoff. At this point, all remaining cross-loadings were $<\!|0.15|$, communalities $>\!0.3$, and complexities $<\!1.1$. We removed one more item (``On the [Physics 1] midterm exam, I felt confident about my performance while taking the solo part of the exam'') because it displayed the lowest primary loading and communality, it is theoretically different than the other self-efficacy items in that it asks about a past test, and removing this item did not significantly change the loadings of the remaining items. This process removed most of the items additional to those directly adapted from the MSLQ, leaving a \textit{self-efficacy} factor containing 9 items and a \textit{test anxiety} factor containing 4 items. 

The self-efficacy items remaining overlapped those of the MSLQ except for one item---which originally referred to ``assignments and tests''---that was split into separate items for ``homework'' and ``tests''. For the self-efficacy factor ($R^2\!=\!.96$), the three highest-loading items---all with primary loading of $0.88$---were ``Considering the difficulty of this course, the instructor, and my skills, I think I will do well in this course,'' ``I'm confident I can master the skills being taught in this course,'' and ``I'm confident I can do an excellent job on tests in this course.'' For the test anxiety factor ($R^2\!=\!.83$), two items were retained from the MSLQ, one retained item was a combination of two MSLQ items (``I feel my heart beating fast, an uneasy feeling in my stomach or tightness in my chest when taking an exam''), and one item was new but loaded well onto this factor (``I get so nervous during a test that I cannot recall the material I have learned''). The latter two items were the highest-loading items, with primary loadings of $0.76$ and $0.80$, respectively.
The final EFA explained $64\%$ of the variance in the included statements. For each student, we computed normalized factor scores using the `tenBerge' correlation-preserving regression method.

We observed significant correlations between the factor scores for self-efficacy and test anxiety and the assessment scores. Self-efficacy was significantly correlated with FCI post-test, $r(385)\!=\!.42$, $p\!<\!.001$, and the exam, $r(436)\!=\!.41$, $p\!<\!.001$. Test anxiety was significantly correlated with FCI post-test, $r(385)\!=\!-.24$, $p\!<\!.001$, and exam, $r(436)\!=\!-.24$, $p\!<\!.001$, though with a smaller effect size than self-efficacy.

\section{\label{sec:results}Analysis and results}

\subsection{\label{sec:parity}Equity of parity in assessments and affective factors}
In our study, gender equity of parity would mean that women and men have comparable distributions in scores for the exam, FCI, and affective measures. As shown in Table~\ref{tab:diffs}, differences can be seen in all our measures, with women having lower scores on measures associated with success (self-efficacy and assessment scores) and a higher score on a possibly detrimental affective factor (test anxiety).

\subsection{\label{sec:fairness}Equity of fairness in FCI post-test and exam scores}

In our study, we interpret gender equity of fairness to mean that a student with the same FCI pretest, self-efficacy, and test anxiety would achieve the same FCI post-test or exam score independent of gender. To examine equity of fairness, we used model-selection criteria to compare different multiple regression models predicting standardized FCI post-test or exam. For the analysis, we consider gender, FCI pretest, self-efficacy, and test anxiety to be possible independent variables. Since a one-way ANOVA showed possible small effects of lecture section on FCI post-test and exam, we considered the inclusion of section as a random variable in the analysis. However, we found that the intraclass correlation $\rho$ was less than $0.05$ for both FCI post-test and exam models; therefore, including section as a random effect would not be expected to significantly improve our models \cite{Theobald2018} and we do not include it.

Starting from the base model with gender predicting FCI post-test or exam score, we added predictors (FCI pretest, self-efficacy, and test anxiety) to the model, using the Akaike information criterion (AIC) to determine if each successive model was a better fit to the data. Motivated by previous work which identified test anxiety as an issue for female students only \cite{Ballen2017a}, we also considered gender-self-efficacy and gender-test anxiety interaction terms. In comparing models, we consider models with $\Delta\textrm{AIC}\!<\!2$ to be equivalent, and look for the simplest model with the most explanatory power. 

For the models predicting exam score, a subset of the models considered are presented in Table~\ref{tab:models}. In the analysis, the categorical variable gender is coded as $\textrm{F}\!=\!1$, $\textrm{M}\!=\!0$. In the base model of gender predicting exam score, $\beta_\textrm{gender}\!=\!-0.40$ implies that a female student would score $0.40$ standard deviations lower than a male student on the exam. However, $R^2\!=\!.03$ for this model, indicating that only 3\% of the variance in the exam scores is explained by gender alone; this base model has little practical explanatory power, and we are led to add more predictors. In considering all combinations of the predictors, including gender-self-efficacy and gender-test anxiety interaction terms, no other model had a lower AIC (within $|\Delta\textrm{AIC}|\!<\!2$) than the model with gender, FCI pretest, and self-efficacy as predictors. In general, including test anxiety as a predictor made models more complex but did not improve the models as much as including self-efficacy. Relative to the model with gender and FCI pretest, the inclusion of self-efficacy explained 7\% more of the variance in the exam data. Compared to the model with gender as the only predictor, the addition of FCI pretest and self-efficacy caused the regression coefficient for gender to change signs, going from one that predicted a statistically significant disadvantage for women to a statistically significant advantage (of $0.20$ standard deviations). Therefore, after taking into account FCI pretest and self-efficacy levels, exam scores may not demonstrate equity of fairness.

The outcome of the model-selection procedure for the FCI post-test regression followed similar trends to that for exam score. Here, the simplest and best-fitting model had only FCI pretest and self-efficacy as predictors---including gender or test anxiety as predictors resulted in a more complex model, but was within $\Delta\textrm{AIC}\!<\!2$ of the simpler model. In this best-fitting model, the coefficients were $\beta_\textrm{FCI pretest}\!=\!0.57$ ($p\!<\!.001$) and $\beta_\textrm{self-efficacy}\!=\!0.18$ ($p\!<\!.001$). For this model, $R^2=.45$, indicating that FCI pretest and self-efficacy explain 45\% of the variance in FCI post-test scores. However, the increase in $R^2$ relative to the model with gender and FCI pretest is $.02$, indicating that self-efficacy carries less additional explanatory power for the FCI post-test than for the exam. The absence of gender as a predictor in the best-fitting model for FCI post-test scores could be interpreted as equity of fairness in FCI post-test achievement. 

\begin{table}[t]
    \begin{ruledtabular}
    \begin{tabular}{ccccccc}
        $\beta_\textrm{gender}$ & $\beta_\textrm{FCI pretest}$ & $\beta_\textrm{self-eff.}$ & $\beta_\textrm{test anx.}$ & $R^2$ & AIC\\\hline
        -0.40*** & & & & .03*** & 1153.1 \\
        0.09 & 0.52*** & & &  .26*** & 1042.2 \\
        0.13 & 0.49*** & & -0.15*** & .29*** & 1031.4\\
        0.20* & 0.43*** & 0.29*** & & .33*** & 1003.6 \\
        0.21* & 0.43*** & 0.26*** & -0.08 & .34*** & 1002.2\\
    \end{tabular}
    \end{ruledtabular}
    \caption{\label{tab:models}Model coefficients and goodness of fit measures for multiple regression models predicting the exam score. Each row corresponds to a different model; if a row is missing a coefficient, that variable was not included in the model. The categorical variable gender is coded as $\textrm{F}\!=\!1$, $\textrm{M}\!=\!0$; all other measures are standardized. The models presented here are illustrative but not fully representative of the complete model selection process described in the text. *: $p\!<\!.05$; **: $p\!<\!.01$; ***: $p\!<\!.001$.}
\end{table}

\section{\label{sec:conc}Discussion}
Consistent with previous reports, we do not observe equity of parity between women and men on our FCI post-test or exam scores. Among the negative effects of this inequity is that these outcomes influence how a person moves forward from this course, whether into further physics courses or along whatever path they choose. For this class, we also do not observe equity of parity in our two affective measures: self-efficacy and test anxiety.

From an equity of fairness perspective, the results suggest a different interpretation. For a given FCI pretest and self-efficacy, female and male students are ending up at the same place for the FCI post-test while women are achieving higher scores for the exam. Therefore, we observe equity of fairness for the FCI post-test but not for the exam; however the inequity in exam results is to the benefit of women, the equity-seeking group. If part of our ultimate goal is to make progress toward equity of parity in assessment results, it must be that equity-seeking groups achieve larger gains, all else equal.

The results presented here represent a pilot study of self-efficacy, test anxiety, and assessment scores in our context. In exploring our data, we made many comparisons and considered many possible models. Thus, it is likely that statistical significance as presented is overstated and we caution against making firm conclusions based on the results. However, the trends shown here are interesting in the context of several recent studies.

In modelling both the FCI post-test and exam, self-efficacy emerged as a significant predictor. The importance of self-efficacy is consistent with several recent studies \cite{Kost-Smith2011,Kalender2020a,Ballen2017}. In Ref. \cite{Ballen2017}, changes in self-efficacy mediated the improvement in assessment scores for the equity-seeking group of under-represented minorities. Since women on average report a lower self-efficacy, this supports the possibility that attending to self-efficacy could specifically benefit women in the course. To better understand if a self-efficacy intervention may work in our context, we could study how changes in self-efficacy relate to assessment results. That self-efficacy held predictive power here is in contrast to a recent report \cite{Salehi2019}. An important difference is that Salehi \emph{et al.} \cite{Salehi2019} found that including a general measure of college preparation (ACT or SAT scores) in addition to a subject-specific preparation gave the best exam-score model. It is possible that our self-efficacy factor may overlap with general college readiness or ability, and that controlling for that aspect would give a different result here. Additionally, general academic preparation has been shown to explain a fraction of the gender difference on concept inventories \cite{Henderson2019}. 

Test anxiety did not show up as a significant predictor in our model-selection process. Women in our study do report higher levels of test anxiety. However, in contrast to Ref. \cite{Ballen2017a}, this does not translate into reduced assessment scores in our data. It could be that women do not actually experience more anxiety during tests even though they report it on the survey; this discrepancy has been demonstrated in a math context \cite{Goetz2013}. Our results echo those of a recent study on a test-anxiety intervention in an introductory biology course, which found that ``women are underperforming in STEM courses for reasons other than ... test anxiety'' \cite{Harris2019a}.

Though we focus our report on the differences between female and male students, our ultimate goal is to alter our classrooms to be more inclusive. Better understanding how students experience our classes affectively, and the relation of these with assessments, offers a promising avenue for suggesting ways to change the culture in our classroom such that all students are welcome and able to realize their potential.

\acknowledgments{E.~Altiere acknowledges support from the UBC Centre for the Integration of Research, Teaching and Learning (CIRTL). We acknowledge the students and the instructors of Physics 1 for their participation.}

\bibliography{Affective_factors_and_gender_in_physics} 

\end{document}